\documentstyle[epsf,twocolumn]{esa_sp_latex}

\begin{document}
%

\parindent 0pt
\parskip 10pt plus 1pt minus 1pt
\hoffset=-1.5truecm
\topmargin=-1.0cm
\textwidth 17.1truecm \columnsep 1truecm \columnseprule 0pt 

\topmargin 1cm

\def\NH{N_{\rm H}}

\title{\bf THE X/GAMMA-RAY SPECTRAL PROPERTIES OF NGC 7172}

\author{{\bf F.~Ryde$^1$, J.~Poutanen$^1$, R.~Svensson$^1$, 
S.~Larsson,$^1$}\\
{\bf 
and S.~Ueno$^2$} \vspace{2mm} \\
$^1$Stockholm Observatory, Saltsj\"obaden, Sweden\\[1.5mm]
$^2$Kyoto University, Sakyo-ku Kyoto, Japan}
\maketitle

\begin{abstract}

We present a combined, non-simultaneous, {\it ASCA} GIS and {\it CGRO} OSSE 
spectrum  of the Seyfert 2 galaxy, NGC 7172, and make broad band spectral 
fits. 
The only Seyfert 2 galaxy  previously studied over such a broad band is
NGC 4945. We find that the most probable model for the data is 
a power law  with an exponential cut-off being affected by a neutral
 absorber. 
The best fit parameters are found to be $\Gamma = 
1.47 \pm 0.15$ and $\NH = (7.8 \pm 0.6) 
\cdot 10^{22}$ cm$^{-2}$. 
The spectral index of the underlying power law of NGC 7172 has therefore
varied from $1.8$ to $1.5$ since the {\it Ginga} observations in 1989. 
For this simple model the e-folding energy at  $88 ^{+65} _{-28}$ keV
is relatively well constrained.
The observed  flux in the $2-10$ keV range is $F_{2-10} = 
4.7 \cdot 10^{-11}$ erg cm$^{-2}$ s$^{-1}$, which corresponds to a 
small increase since the {\it{Ginga}} measurement in October 1989.

Keywords: Galaxies: individual: NGC 7172; Galaxies: Seyfert; X-rays: 
galaxies.

\end{abstract}

\section{INTRODUCTION}

In the unified model of Seyfert galaxies (\cite{antonucci}), the
observed differences between Seyfert 1 and Seyfert 2 galaxies are
explained as an orientation effect caused by an obscuring molecular
torus that surrounds the central X-ray source. This intrinsic X-ray source
is suggested to be the same for all Seyfert galaxies.
Some Seyfert 2 galaxies have been observed to have an intrinsic spectrum
that is flatter than what has been observed for Seyfert 1's
(\cite{SmithDone}), which contradicts the unified model. However, when
observing with a single satellite, normally only a narrow spectral band
can be studied (e.g., $0.5 - 12$ keV for {\it ASCA}). If the observed 
object is heavily absorbed, this range could be too small to be 
able to determine whether the 
intrinsic spectrum really is flat or whether it is intrinsically steep 
and maybe affected by Compton reflection. This may not be the case
for broad band spectra obtained by combining data from two or more 
satellites. In this paper, we conduct such an analysis on
the Seyfert 2 galaxy NGC 7172. 
The only Seyfert 2 galaxy previously 
studied over a broader band using non-simultaneous 
{\it ASCA}/{\it Ginga}/{\it CGRO} OSSE data
is NGC 4945 (\cite{MadDone}). 
The average non-simultaneous {\it Ginga}/{\it CGRO} OSSE spectrum 
of three Seyfert 2 galaxies has also been studied (\cite{Z5}).

The X-ray spectrum of NGC 7172, an obscured, edge-on Seyfert 2 galaxy  
at a redshift of $z = 0.008$ (\cite{Sharples}),
was studied 
by {\it EXOSAT} in October 1985 (\cite{TurnerPounds}),
by {\it Ginga} in October 1989 (\cite{Warwick}; \cite{NandraPounds};
\cite{SmithDone}), by {\it CGRO} OSSE in March 1995, by {\it ASCA} 
on May 13, 1995 (\cite{Ryde}), and on
May 17, 1996 (\cite{Matt}). Here we analyse and fit the March/May 1995 data.
The {\it ASCA} data for NGC 7172
are extracted using XSELECT (ftools v3.6) and all the data analysis is
performed with XSPEC.

\section{THE X/$\gamma$ OBSERVATIONS OF NGC 7172}

We observed NGC 7172 with the {\it ASCA} satellite 
from 23:18 UTC on May 12
until 16:00 UTC on May 13, 1995 with the two GSPC's, GIS2 and GIS3 
and with the two CCD detectors SIS0 and SIS1. 
The SIS data seem to be sensitive to the screening criteria and give 
different results depending on the criteria adopted. The SIS0 and the SIS1 
detectors also give different results as compared to each other. This needs 
to be investigated further. The GIS data are, however, consistent with
 each other and  they are usually
less affected by calibration problems. In this paper we therefore only
use the GIS2 and the GIS3 data for the {\it ASCA} observation. 
In screening the GIS data, standard, conservative criteria are applied: 
Minimum angle to Earth limb $= 10^\circ$, angle 
from bright Earth (illuminated limb) $ = 20^\circ$, minimum cutoff
rigidity $ = 6$ GeV/c. 
The background is extracted from a source-free
region of the same image as the source. The final results do not differ
significantly, if a background instead is extracted from
the blank sky background event files provided by the  {\it ASCA} Guest
Observer Facility. The extracted data are regrouped to contain a minimum 
of $20$ counts per channel to allow the use of the $\chi^2$-test. 
This leads us to ignore low energy channels up to $\sim 2$ keV. 
The net exposure time for both detectors is $27$ ks.
 NGC 7172 did not show any significant variability during the observations,
with the variations being less than 10 per cent of the mean count rate.

 {\it CGRO} OSSE observed NGC 7172 from February 21 to March 21, 1995.
 The data may be affected by the proximity of PKS 2155-304, as both objects 
often had to be observed together. The OSSE data were provided by
 the CGRO Science Support Center.

\section{DATA ANALYSIS AND RESULTS}

We analyse the combined spectrum using the
{\it ASCA} GIS, and {\it CGRO} OSSE data discussed above
 from which one should be able to determine 
the photon index unambiguously. The observations are not
simultaneous, however, but the {\it ASCA} and OSSE observations 
are only two months apart. 

We start off with a model (denoted as `model P') consisting of a power
 law and a 
neutral absorber, leaving the e-folding  energy free.
The resulting fit gives a photon spectral index $\Gamma =  1.47 \pm 0.15$,
an e-folding energy of 
$88 ^{+65} _{-28}$ keV and an equivalent hydrogen column density $\NH = 
(7.8 \pm 0.6) \cdot 10^{22}$ cm$^{-2}$. The fit has a reduced $\chi ^2$
$ = 0.881$, for the number of degrees of freedom, dof $=470$.
 For the simple model under consideration, the OSSE data
gives a relatively good constraint to the e-folding energy.
The observed flux, in the $2-10$ keV range, $F_{2-10} =  4.7 \cdot 
10^{-11}$ erg cm$^{-2}$ s$^{-1}$ and the flux, corrected for
 absorption, $F_{2-10}^{\rm no-abs} = 7.3 \cdot 10^{-11}$ erg cm$^{-2}$ 
s$^{-1}$. The luminosity, with $z = 0.008$, $H_{\circ} = 50$ km s$^{-1}$
Mpc$^{-1}$, $ q_{\circ} = 0$, becomes $L_{2-10} = 1.5 \cdot 10^{43}$ 
erg s$^{-1}$.

In the next step of the analysis, we try adding a Compton reflection 
component (XSPEC model {\tt{pexrav}}  by \cite{magdziarz}).
The resulting fit does not change, and the
reflection parameter stays at zero. The data clearly do not require a 
reflection component and we therefore use `model P' above as our best fit
model.

The {\it Ginga} observations of NGC 7172 from 1989 (\cite{Warwick}) 
showed that $\Gamma = 1.85 ^{+0.05} _{-0.04}$ and the $ F_{2-10} = 4.0 
\cdot 10^{-11}$ erg cm$^{-2}$ s$^{-1}$.
The X-ray spectrum of NGC 7172 has thus become flatter between 1989
and 1995, and the $2-10$ keV flux level has increased.

\begin{figure}[htbp]
  \begin{center}
    \leavevmode
\epsfxsize=8.0cm
\epsfbox[74 360 488 720]{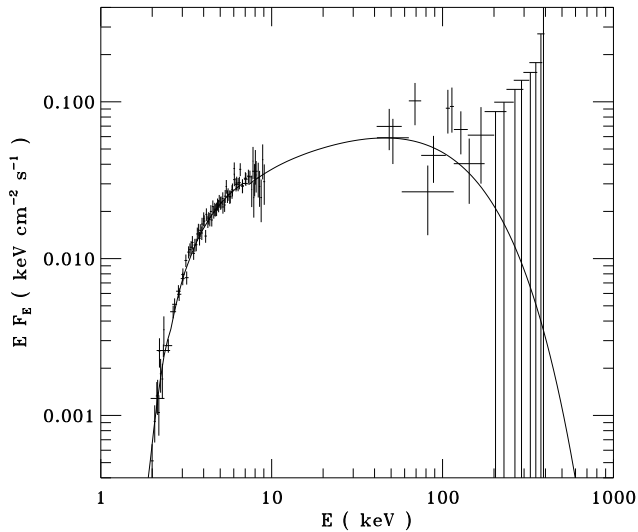}
  \end{center}
  \caption{\em A $\nu F _\nu$ plot of the combined {\it ASCA}/{\it
  CGRO} OSSE data of NGC 7172 together with the best fit.}
 \label{fig.2}
\end{figure}

Figure \ref{fig.2} shows the combined  {\it ASCA} GIS and {\it CGRO}
 OSSE data and the best
fit model (`model P') in $E \cdot F _E$ versus
$E$. The  {\it ASCA} data below $\sim 2$ keV do not have sufficient 
photon counts for a proper $\chi ^2$ analysis and that data was not used 
in the modeling. It is, however, evident that there
exists a soft component, which is often seen in Seyfert
2 galaxies ({\cite{Morse}}).

\section{SUMMARY}

Using the {\it ASCA} GIS and the {\it CGRO} OSSE  observations of
 NGC 7172, a broad  
spectral band is achieved. The best fit model is  
a power law affected by a neutral absorber, $\Gamma =  1.47 \pm 0.15$ and
 $\NH = (7.8 \pm 0.6) \cdot 10^{22}$ cm$^{-2}$. The e-folding energy
 is  constrained by the OSSE data to be 
$E _c = 88 ^{+65} _{-28}$ keV.  
The fit does not improve, when more complicated
 models are used, such as adding a reflection component.
 The flat spectrum found means
that the spectral index of NGC 7172 is variable, changing from
$\Gamma = 1.8$ in 1989 to $\Gamma = 1.5$ in 1995. It still 
lies within the range  that Seyfert  galaxies are observed to have (see, 
e.g. \cite{NandraPounds}). We note that some other Seyfert
galaxies, most notably NGC 4151 (\cite{YW}), also have variable
$\Gamma$.

\section{ACKNOWLEDGMENTS}

We thank Drs. Christine Done, Matteo Guainazzi and  Giorgio Matt
 for enlightening discussions. 
We are also indebted to Dr. Tom Bridgman at the CGRO Science Support
 Center and  to the  {\it ASCA} GOF at Goddard Space Flight Center 
for their assistance. 
This research was supported by the Swedish National
Space Board and the Swedish Natural Science Research Council.

\end{document}